\documentclass[aps,preprint,fleqn,prb,byrevtex,citeautoscript,showpacs,superscriptaddress]{revtex4}

\usepackage{psfrag,graphicx,longtable}
\newcommand{\Bra}[1]{\left \langle #1 \right |} %<1|
\newcommand{\Ket}[1]{\left | #1 \right \rangle}  %|2>
\newcommand{\ME}[3]{\Bra{#1} #2 \Ket{#3}} %  <1|2|3>
\begin{document}  

\title{
Local Field effects on the radiative lifetime of emitters in surrounding media:
virtual- or real-cavity model?
}
\author{Chang-Kui Duan}
\affiliation{
Institute of Modern Physics,
Chongqing University of Post and Telecommunications, Chongqing 400065, China.
}\author{Michael F. Reid}
\affiliation{
Department of Physics and Astronomy and MacDiarmid Institute of
Advanced Materials and Nanotechnology, University of Canterbury, Christchurch,
  New Zealand}
\author{Zhongqing Wang}
\affiliation{
Institute of Modern Physics,
Chongqing University of Post and Telecommunications, Chongqing 400065, China.
}

\begin{abstract}
For emitters embedded in media of various refractive indices,
 different macroscopic or microscopic theoretical models
 predict different dependencies of the spontaneous emission
 lifetime on refractive index. Among those models are the two
most promising models: the virtual-cavity
 model and the real-cavity model. It is a priori not clear 
which model is more relevant for a given situation. 
By close analysis of the available experimental results and examining the 
assumptions underlying the two models, we reach a consistent interpretation
of the experimental results and give the criteria which model should
apply for a given situation.
\vskip 0.5cm


\end{abstract}

\date{19th  May, 2005}
%\noindent Contact Details:\\
%Chang-Kui Duan\\
%Institute of Modern Physics,\\
%Chongqing University of Post and Telecommunications,\\
%Chongqing 400065,  P. R. China\\
%%Email: chang-kui.duan@canterbury.ac.nz; duanck@cqupt.edu.cn\\
%Phone: +86 (23) 62471434, Fax: +86 (23) 62471434
%\vskip 2cm

\maketitle

It has long been known that the spontaneous-emission rates of emitters 
can be modified by changing the surrounding dielectric media.\cite{Blo1965,Top2003}
 The theory on this subject continues to attract considerable attentions due to 
its fundamental importance and its relevance to various
applications in low-dimensional optical materials and photonic
crystals\cite{Luk2002,Kum2003,Ber2004}. Various
macroscopic (see Ref. \onlinecite{Top2003} for a recent review) and
microscopic \cite{Bor1999,Cre2000,Ber2004} theoretical models have
been developed to model the dependence of the
spontaneous emission rates (or lifetimes) on refractive index. 
Among those models are the real-cavity model (also referred
to as empty-cavity model\cite{Top2003}), where emitters (usually
ions) are assumed to create tiny cavities when replacing
host ions, and the virtual-cavity model, which is based on the 
Lorentz local field\cite{Bor1999}.
Different models predict substantially different 
dependences of radiative lifetime on the surrounding media.
There are also some measurements intended to discriminate these models,
with most experimental results tend to agree with the real-cavity model.
Especially, recent measurements on the radiative lifetime of Eu$^{3+}$ ion
embedded in glass of varying refractive index\cite{Kum2003} also tend to agree with 
the real-cavity model, in contrary to the general belief that the virtual-cavity
model should be more relevant. Duan and Reid \cite{Dua2005c}
pointed out that $4f\rightarrow 4f$ electric dipole radiative relaxation,
which is due to mixing in $4f^N$ states with states of opposite parity,
depends strongly on the environment and does not 
serve as a good examination of the two models. To overcome the problem in
$4f\rightarrow 4f$ transitions, Duan and Reid\cite{Dua2005c} analyzed
the lifetimes of $5d\rightarrow 4f$ transitions
of Ce$^{3+}$ ions in hosts of different refractive indices and the results
maintained the text-book virtual-cavity model. However, they did not try to answer
the questions why most other experimental results agree with the real-cavity 
model, and which model should apply for a given situation. 

In this Letter, we examine those reported experimental results on various
emitters in various surrounding media that appear to support different models.
We point out why they support certain model in some cases and why they
need to be reinterpreted with other models. Finally we answer the titled
question by our consistent explanation of the experimental results and
state the rule to choose the proper model for a given situation.

\begin{figure}[htp]
\includegraphics[width = 12cm] {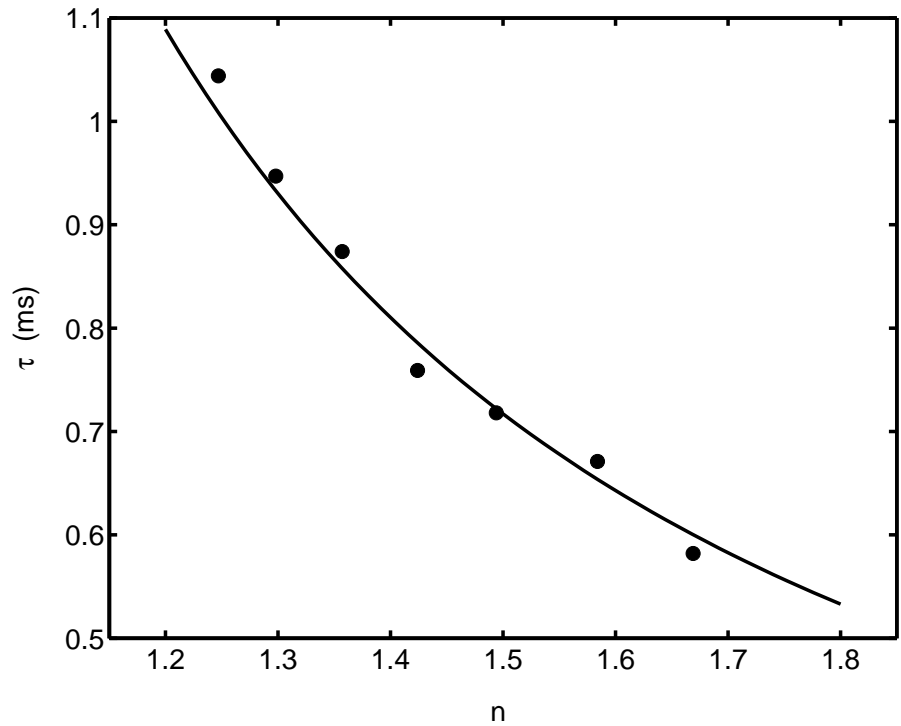}
\caption{
\label{eu-hfa-topo}
Radiative lifetime of Eu$^{3+}$-hfa-topo as a function of the solvent refractive
index. The dots are measured lifetimes and the curve is fit to Eq.(\ref{real})}
\end{figure}

{\bf Measurements on the radiative 
lifetime of Eu$^{3+}$-hfa-topo complex emitter} in a
 series of apolar hydrocarbons was reported by Rikken {\it et al.} in
1995\cite{Rik1995}. Experimental results show that the emissions are
from ${}^5D_0$ levels and dominated by electric dipole transitions to
${}^7F_2$ levels, and the radiative quantum yield is close to unity.
Fig. \ref{eu-hfa-topo} is the plot of the measured lifetimes
together with the best fitting using the real-cavity model:
\begin{equation}
\label{real}
\tau_{\rm real} = \tau_0 \frac{1}{n \left ( \frac{3n^2}{2n^2+1} \right )^2},
\end{equation}
where $\tau_0$ is the lifetime of the emitter in vacuum, the factor $n$ is due to
 the reduction of speed of light in the media and
$\left ( \frac{3n^2} {2n^2+1} \right )^2$ is the ratio of the electric field in the
cavity (local field) to the macroscopic field in the media.  $\tau_0$ is taken to be
an adjustable parameter independent of the refractive index in the fitting.

%
%figure1
%\input{figure1.tex}

It is well-known from Fermi's golden rule that  $\tau_0$ is proportional to 
the square modulus of the electric dipole moment and the cubic of the emission photon energy.
In general, the $4f-4f$ emission photon energy does not change substantially in different
surrounding media, but the electric dipole moment of $4f-4f$ transitions depends strongly 
on the arrangement of ligand ions. In the experiment, Eu$^{3+}$ ions are embedded in organic ligand cages
which remain almost the same when the surrounding media are changed. Therefore the electric dipole
moments and emission photon energies do not vary, and hence the unknown $\tau_0$ does not vary with
the surrounding media and can be treated as an adjustable parameter.

It can be explained why the radiative lifetime of Eu$^{3+}$-hfa-topo complex
supports the real-cavity model. The complex expels the solvent 
from the volume occupied by the complex and create a real tiny cavity of the dielectric, 
and the relatively large and rigid Eu$^{3+}$ complex leads to a region of small polarizability 
and hence small refractive index, which in the limiting case, is close to that of the vacuum.
The ratio of the average electric field of the complex to the macroscopic electric field
in the surrounding media obeys the real-cavity model. The ratio of the  electric field ``felt'' by
the Eu$^{3+}$ ion to the average electric field of the complex is a constant that does not
depend on the surrounding media. Therefore the radiative lifetime for a given complex in various
surrounding media can be described with the real-cavity model.

The real-cavity model is also confirmed by the measurements on Eu$^{3+}$(fod)$_3$  embedded in a dense
supercritical CO$_2$ gas, where the refractive index changes from 1.00 to 1.27 when the pressure
varies between 1 and 1000 bars. 

Other similar cases have been reported by Lavallard {\it et al.} in 1996\cite{Lav1996}
and later by Lamouche {\it et al.} in 1998\cite{Lam1998} with the sulforhodamine
 B molecules as emitters. In those experiments,
the sulforhodamine B molecules are dissolved in water droplets (with refractive index $n_{\rm w}$) 
which are stabilized by a surfactant (with refractive index $n_{\rm s}$) and then suspended
in solutions of different refractive index $n$. Theoretical modeling for this case is 
 slightly more complicated and more general than the real-cavity model. It has been given 
based on the same principle as the real-cavity model as:\cite{Lav1996}
%\begin{widetext}
\begin{eqnarray}
\label{realb}
&&\tau_{\rm real} = \tau_0 \frac{1}{nB^2},\\
\nonumber
&&B = \frac{9n^2 n_{\rm s}^2}{(n_{\rm s}^2+2n^2) (n_{\rm w}^2+ 2 n_{\rm s}^2) + 
            2 (\frac{b}{a})^3 (n_{\rm w}^2 - n_{\rm s}^2) (n_{\rm s}^2 - n^2)},
\end{eqnarray}
%\end{widetext}
where $a$ is the total radius and $b$ is the radius of the water core. The special cases 
where $n_{\rm w} \approx n_{\rm s}$ and $a \approx b$ reduce to the real-cavity model 
(by replacing the $n$ in Eq.(\ref{real}) with $n_r = n/n_{\rm w}$). The measurements confirm 
the more general equations  Eq.(\ref{realb}) than the real-cavity model Eq.(\ref{real}). 

\begin{figure}[htp]
\includegraphics[width = 12cm]{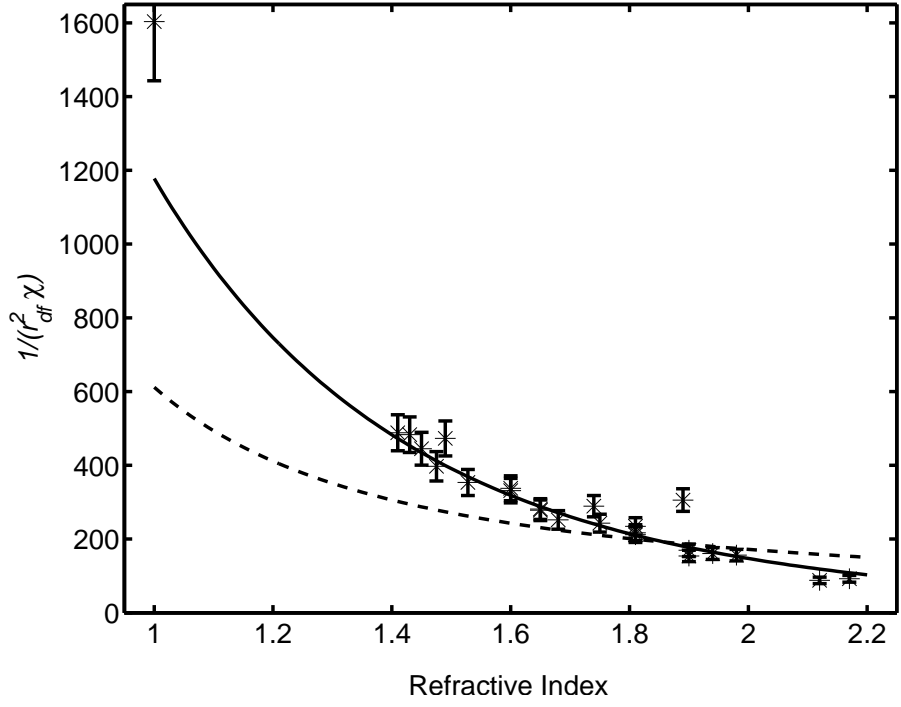}
\caption{
Variation of $1/(|\ME {5d} r {4f}_{\rm eff}|^2 \chi)$ with refractive
index. The experimental values are plotted as `*' with a $10\%$ error
bar. The solid curve is calculated with virtual-cavity model using
best least-square-fitting value $\ME {5d} r {4f}_{\rm eff} =  0.0291$, 
 and the dashed curve is for real-cavity model with best least-square-fitting
value $\ME {5d} r {4f}_{\rm eff}^{\prime} = 0.0404$.
\label{ceiv}
}
\end{figure}

{\bf Measurements on the radiative lifetime of Ce$^{3+}$ ions} in hosts of different 
refractive indices have been summarized by Duan and Reid\cite{Dua2005c}. The transition
involved is the $5d\rightarrow 4f$ electric dipole allowed emission whose electric dipole 
moment is proportional to  the radial integral $\ME{5d}{r}{4f}$, which is mainly determined
by the localized orbitals of Ce$^{3+}$.
Assuming $\ME{5d}{r}{4f}$ to be constant, the results do not agree with the
real-cavity model and are much better described by the virtual-cavity model,
{\it i.e.} 
\begin{eqnarray}
1/\tau_{\rm virtual}&=& \frac{64\pi^4 e^2 |\ME {5d}{r}{4f}|^2 }{5h \lambda^3} 
             \chi\\
&=& \frac{4.34 \times 10^{-4} |\ME {5d}{r}{4f}|^2 }{\lambda^3}
             \chi,              
\end{eqnarray}
where  $\lambda$ is the the average emission wavelength
for Ce$^{3+}$ in a given host, and
\begin{equation}
\chi = n \left ( \frac{n^2+2}{3}\right )^2
\end{equation}
is the refractive-index dependent factor.
 Note that $\lambda$ can be substantially 
different for Ce$^{3+}$ ions in different hosts. From the measured lifetime $\tau_{\rm exp}$ and emission
wavelength $\lambda$, we can obtained ``experimental'' values
\begin{equation}
\frac{1}{[|\ME {5d}{r}{4f}|^2 \chi]_{\rm exp}} = \frac{4.34\times 10^{-4}\tau_{\rm exp}}{\lambda^3}.
\end{equation}
Those ``experimental'' values can be compared to calculated values by considering the
the unknown $|\ME {5d}{r}{4f}$ as a fitting parameter. In Fig.\ref{ceiv},
the ``experimental'' values $1/[|\ME {5d}{r}{4f}|^2 \chi]_{\rm exp}$ are
plot as a function of $n$ together with the best least square fitting using
the virtual-cavity model ($\chi = n [(n^2+2)/3]^2$). For comparison,
the least square fitting using the real-cavity model ($\chi = n [3n^2/(2n^2+1)]^2$)
is also plot. 

On the other hand, we can calculate $\ME {5d}{r}{4f}$ for each Ce$^{3+}$ in each hosts
for both virtual- and real-cavity models. For the virtual-cavity model, the obtained values
show no systematic dependence on refractive index $n$ as expected, but there is 
for the real-cavity model. This also means that the virtual-cavity model describes 
the radiative lifetime of Ce$^{3+}$ in various hosts better than the real-cavity model.

%figure2

%\input{figure2.tex}
%

The results can also be understood. In all those hosts, the low-polarizability ions 
Ce$^{3+}$ replace cations with low polarizability. These replacements do not create 
(tiny) cavities of the dielectric 
media, substantially different from the real-cavity cases. Hence the ratio
of the local field to the macroscopic field can be calculated 
using the Lorenz model, {\it i.e.} the virtual-cavity model.

\begin{figure}[htp]
\includegraphics[width = 12cm]{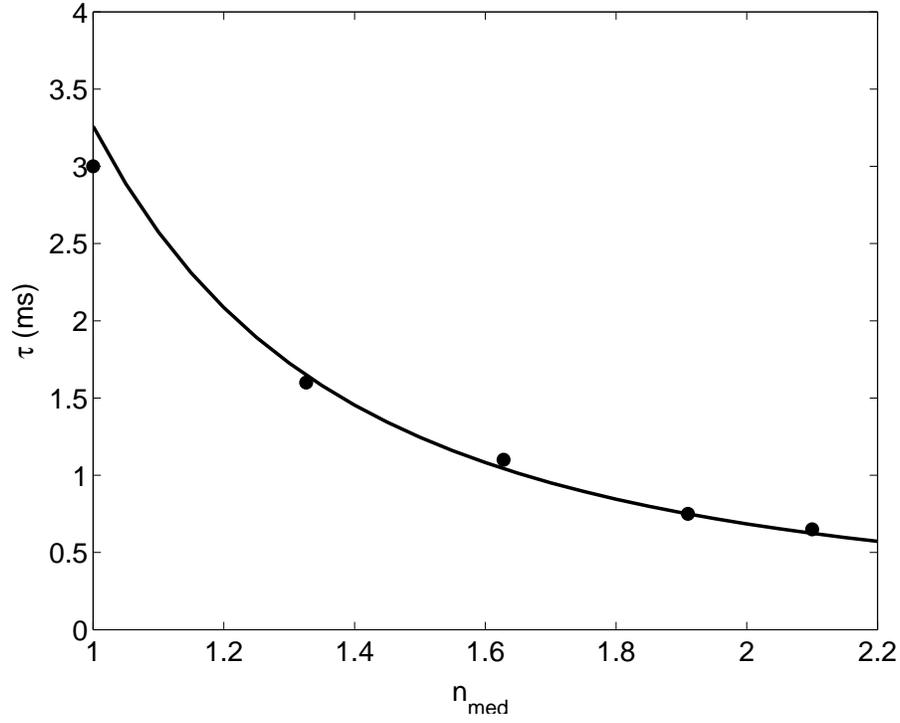}
\caption{
\label{euy2o3}
The dependence of the ${}^5D_0$ radiative lifetime $\tau_{\rm R}$ for the Eu$^{3+}$ 
C site on the refractive index of the media $n_{\rm med}$. 
Solid line: simulation with Eq. (\ref{newmodel}) using $x=0.15$; 
dots: experimental values as given in FIG. 2 of Ref. \onlinecite{Mel1999}.
}
\end{figure}

{\bf The radiative lifetime of Y$_2$O$_3$:Eu$^{3+}$ nanoparticles 
embedded in media of various refractive indices}
were measured in 1999\cite{Mel1999}. 
The experimental results were originally interpreted with the virtual cavity model.
Since the nanoparticles (with constant refractive index $n_{\rm Y_2O_3}$) expel,
as has been pointed out by Duan and Reid\cite{Dua2005d}, 
the media (with effective refractive index $n_{\rm eff}(x)$) and create 
real cavities in the media, the real-cavity model should apply instead, with the refractive
index in Eq.(\ref{real}) replaced with the effective refractive index $n_r$ for 
the media relative to the nanoparticle:
\begin{eqnarray}
n_r&=& n_{\rm eff}(x)/n_{\rm Y_2O_3}\\
   &=& \frac{\cdot n_{\rm Y_2O_3} + (1-x)\cdot n_{\rm med}}{n_{\rm Y_2O_3}},
\end{eqnarray}
where $n_{\rm med}$ is refractive index of the media without embedding Y$_2$O$_3$ particles,
and $x$ is the 'filling factor' showing what fraction of space is occupied by the 
Y$_2$O$_3$ nanoparticles surrounded by the media.
At the same time, the $\tau_0$ should be replaced with the lifetime of a Y$_2$O$_3$:Eu$^{3+}$ nanoparticle
in the media such that $n_r =1$. One such case is when the Y$_2$O$_3$:Eu$^{3+}$ nanoparticle
is part of the bulk Y$_2$O$_3$:Eu$^{3+}$, whose lifetime is denoted as $\tau_{\rm bulk}$.
The variation of the lifetime with $n_r$ can be written as:
\begin{eqnarray}
\label{newmodel}
&&\tau_{\rm R} = \frac{1}{\Gamma_{\rm R}}=\tau_{\rm bulk}\frac{1}{n_r} (\frac{2n_r^2+1}{3n_r^2})^2.
\end{eqnarray}
Fig. \ref{euy2o3} plots the radiative lifetime as a function of the media.
It can be seen that Eq. (\ref{newmodel}) with filling factor $x=0.15$ fits
the measurements very well. The filling factor is not explicitly measured, but
this value is consistent with the TEM pictures of these samples\cite{Mel1999}. 

%figure3
%\input{figure3.tex}

\begin{figure}[htp]
\includegraphics[width=12cm]{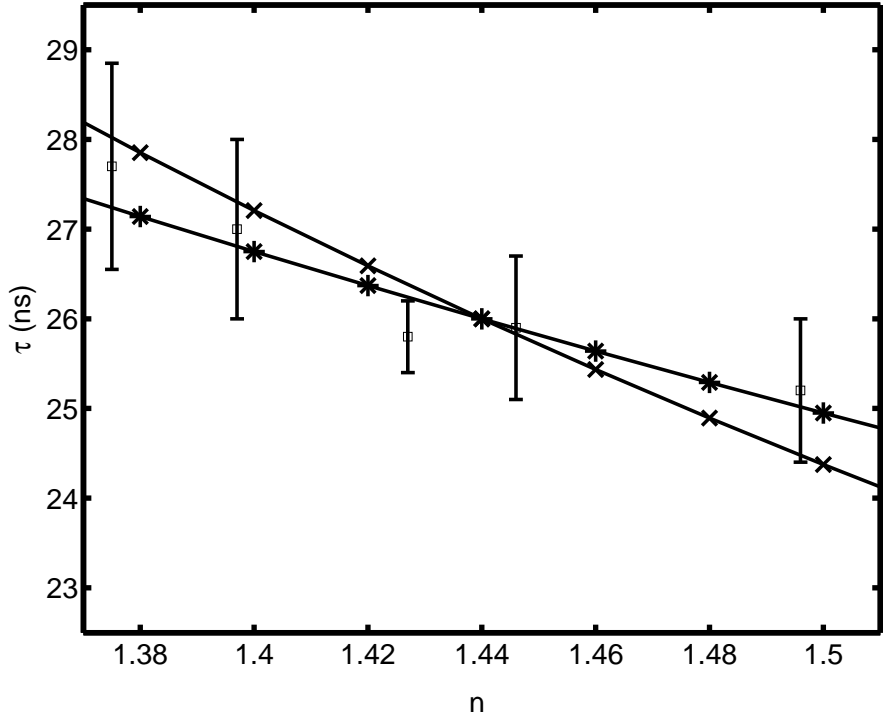}
\caption{
\label{quantumdot}
The dependence of the lifetime of excitons in GdTe quantum dots 
on the refractive index of solvent.
Squares: experimental data taken from FIG. 4 of Ref. \onlinecite{Wui2004};
stars and crosses: simulated with real-cavity model using best-fitted $\tau_{\rm NR} = 58 {\rm ns}$
(giving the experimental quantum efficiency $\sim 55\%$ at $n=1.44$) and a much higher
$\tau_{\rm NR} = 260 {\rm ns}$ (giving a quantum efficiency $\sim 90\%$ at $n=1.44$), respectively.
}
\end{figure}

{\bf The lifetimes of CdTe and CdSe quantum dots} in media\cite{Wui2004} have
been measured very recently and interpreted with the fully microscopic
model of Crenshaw and Bowden\cite{Cre2000}. This model predicts a much weaker dependence
of radiative lifetime on refractive index and does not separate the coefficients
due to photon density of states and due to local-field effect.
More recently, Berman and Milonni\cite{Ber2004} studied the corrections
 to radiative lifetime theoretically
using a slightly different but more realistic approach of fully microscopic many-body theory
 and obtained a theoretical result compatible with macroscopic models. They also pointed
out the problem with the fully microscopic model given by Crenshaw and Bowden.
As pointed out by Duan and Reid\cite{Dua2005b},
since the media are expelled from the space occupied by the quantum dots, 
the real-cavity model should apply. The relaxation in the quantum dots is partly due to 
nonradiative relaxation, which shall not depend on the refractive index 
of the media. We assume that the nonradiative
relaxation rate $1/\tau_{\rm NR}$ is constant and obtain:
\begin{eqnarray}
\frac{1}{\tau(n)} = \frac{1}{\tau_{\rm R0}} (\frac{3n^2}{2n^2+1})^2 n+ \frac{1}{\tau_{\rm NR}}.
\end{eqnarray}
Fig. \ref{quantumdot} plots the total lifetime of the quantum dot CdSe as a function of the
refractive index of the media. Using the experimental quantum efficiency of $55\%$, the
model fits the experimental lifetimes very well. We notice that there is an
 argument in Ref.\ \onlinecite{Wui2004}
that the effective quantum efficiency is higher than $55\%$ and hence plot 
an alternative simulated lifetime using a quantum efficiency $90\%$. It is also 
compatible with the experimental lifetimes within experimental errors. This means that,
although not conclusive, the lifetimes of CdTe and CdSe quantum dots in media
can also be explained by the real-cavity model.

%figure4
%\input{figure4.tex}
%

{\bf The lifetimes of Eu$^{3+}$ ions in the glass system $x$ PbO$+(1-x)$ B$_2$O$_3$}
($x=0.7$-$1.0$)
have been measured in 2003 and interpreted with the real-cavity model.
It has previously been pointed out by Duan and Reid\cite{Dua2005c} that
the electric dipole moment for Eu$^{3+}$ ions $f-f$ transitions strongly
depends on the local structure, which may change when the composition of 
the glass changes, and so the lifetime may not serve as a good examination
of different models. However, they did not intend to explain why the 
experimental results come out to follow the prediction of the real-cavity model.
 Clearly,  if the virtual-cavity model is employed to calculate the calculate the
electric dipole moment, there will be a correlation in the trend of 
the electric dipole moment with the refractive index in that when
the refractive index increases, the electric dipole moment decreases in 
a systematic way.
 Another problem we try to answer is: if Ce$^{3+}$,
whose emission is $f-d$ transition, is used instead of Eu$^{3+}$ in the
glass system, then which model should best describe the measurements.

From a recent study of the network structure of  B$_2$O$_3$-PbO\cite{Ush2004},
it can be seen that the bond length B-O is less than $0.15$ nm,  while 
Pb-O is between $0.25$ to $0.3$ nm, comparable with the usual
bond length between a rare-earth ion and O ion. Therefore a rare-earth ion
can only substitute a Pb ion in the glass. When $x$ decreases from $1.0$ to
$0.7$, the rare-earth ion remains in Pb position and hence the local structure does
not change substantially. Therefore, when $x$ decrease the emission energy does not
change, and it is very likely that the electric dipole line strength of Eu$^{3+}$ 
in the glass do not change substantially either. Hence the variation of experimental 
lifetime with $x$ reflects the effect of changing refractive index, which favor
the real-caivty model.

In factor, We notice that Pb is different from light cations in that it has a very
large polarizability and is often used as additives to glass to increase the index
of refraction. The substitution of Pb with a low-polarizability rare-earth ion
creats a tiny cavity of the dielectric in the media. Hence the real-cavity become
more relevant than the virtual-cavity model. If $f-d$ transition of Ce$^{3+}$ ion
is used instead of $f-f$ transiton of Eu$^{3+}$, we predict that the radiative lifetime
will also close to the real-cavity model and provide an even better test for our
analysis.

In summary, All the experimental results for emitters  embedded in 
dielectric media have been successfully interpreted  
 with either real-cavity model or virtual-cavity model in a consistent way.
 From the interpretation, the titled question is now clear: when the emitter 
expels the dielectric media and create a real (tiny) cavity in the media 
(including the case of ions of high polarizability being replaced with
ionic emitters of low polarizability),
the spontaneous emission lifetime obeys the real-cavity model. 
When the emitters substitute ions
of low polarizability, they do not create a real cavity in the media,
and then the emission lifetime obeys the virtual-cavity model. 

{\bf Acknowledgment} C.K.D. acknowledges support of this work by the
 National Natural Science Foundation of China, Grant No. 10404040 and 10474092.

\newpage

\bibliography{lifetime,celifetime}

\begin{thebibliography}{16}
\expandafter\ifx\csname natexlab\endcsname\relax\def\natexlab#1{#1}\fi
\expandafter\ifx\csname bibnamefont\endcsname\relax
  \def\bibnamefont#1{#1}\fi
\expandafter\ifx\csname bibfnamefont\endcsname\relax
  \def\bibfnamefont#1{#1}\fi
\expandafter\ifx\csname citenamefont\endcsname\relax
  \def\citenamefont#1{#1}\fi
\expandafter\ifx\csname url\endcsname\relax
  \def\url#1{\texttt{#1}}\fi
\expandafter\ifx\csname urlprefix\endcsname\relax\def\urlprefix{URL }\fi
\providecommand{\bibinfo}[2]{#2}
\providecommand{\eprint}[2][]{\url{#2}}

\bibitem[{\citenamefont{Bloembergen}(1965)}]{Blo1965}
\bibinfo{author}{\bibfnamefont{N.}~\bibnamefont{Bloembergen}},
  \emph{\bibinfo{title}{Nonlinear Optics}} (\bibinfo{publisher}{Benjamin},
  \bibinfo{address}{New York}, \bibinfo{year}{1965}).

\bibitem[{\citenamefont{Toptygin}(2003)}]{Top2003}
\bibinfo{author}{\bibfnamefont{D.}~\bibnamefont{Toptygin}},
  \bibinfo{journal}{J. Fluoresc.} \textbf{\bibinfo{volume}{13}},
  \bibinfo{pages}{201} (\bibinfo{year}{2003}).

\bibitem[{\citenamefont{Luks and Perinova}(2002)}]{Luk2002}
\bibinfo{author}{\bibfnamefont{A.}~\bibnamefont{Luks}} \bibnamefont{and}
  \bibinfo{author}{\bibfnamefont{V.}~\bibnamefont{Perinova}}, in
  \emph{\bibinfo{booktitle}{Progress in Optics}}
  (\bibinfo{publisher}{Elsevier}, \bibinfo{address}{Amsterdam},
  \bibinfo{year}{2002}), vol.~\bibinfo{volume}{43}, pp.
  \bibinfo{pages}{295--431}.

\bibitem[{\citenamefont{Kumar and Rao}(2003)}]{Kum2003}
\bibinfo{author}{\bibfnamefont{G.~M.} \bibnamefont{Kumar}} \bibnamefont{and}
  \bibinfo{author}{\bibfnamefont{D.~N.} \bibnamefont{Rao}},
  \bibinfo{journal}{Phys. Rev. Lett.} \textbf{\bibinfo{volume}{91}},
  \bibinfo{pages}{203903} (\bibinfo{year}{2003}).

\bibitem[{\citenamefont{Berman and Milonni}(2004)}]{Ber2004}
\bibinfo{author}{\bibfnamefont{P.~R.} \bibnamefont{Berman}} \bibnamefont{and}
  \bibinfo{author}{\bibfnamefont{P.~W.} \bibnamefont{Milonni}},
  \bibinfo{journal}{Phys. Rev. Lett.} \textbf{\bibinfo{volume}{92}},
  \bibinfo{pages}{053601} (\bibinfo{year}{2004}).

\bibitem[{\citenamefont{Born and Wolf}(1999)}]{Bor1999}
\bibinfo{author}{\bibfnamefont{M.}~\bibnamefont{Born}} \bibnamefont{and}
  \bibinfo{author}{\bibfnamefont{E.}~\bibnamefont{Wolf}},
  \emph{\bibinfo{title}{Principles of Optics}} (\bibinfo{publisher}{Cambridge
  University Press}, \bibinfo{address}{Cambridge}, \bibinfo{year}{1999}).

\bibitem[{\citenamefont{Crenshaw and Bowden}(2000)}]{Cre2000}
\bibinfo{author}{\bibfnamefont{M.~E.} \bibnamefont{Crenshaw}} \bibnamefont{and}
  \bibinfo{author}{\bibfnamefont{C.~M.} \bibnamefont{Bowden}},
  \bibinfo{journal}{Phys. Rev. Lett.} \textbf{\bibinfo{volume}{85}},
  \bibinfo{pages}{1851} (\bibinfo{year}{2000}).

\bibitem[{\citenamefont{Duan and Reid}()}]{Dua2005c}
\bibinfo{author}{\bibfnamefont{C.~K.} \bibnamefont{Duan}} \bibnamefont{and}
  \bibinfo{author}{\bibfnamefont{M.~F.} \bibnamefont{Reid}}, \bibinfo{note}{to
  be published on Current Applied Physics}.

\bibitem[{\citenamefont{Rikken and Kessener}(1995)}]{Rik1995}
\bibinfo{author}{\bibfnamefont{G.~L. J.~A.} \bibnamefont{Rikken}}
  \bibnamefont{and} \bibinfo{author}{\bibfnamefont{Y.~A. R.~R.}
  \bibnamefont{Kessener}}, \bibinfo{journal}{Phys. Rev. Lett}
  \textbf{\bibinfo{volume}{74}}, \bibinfo{pages}{880} (\bibinfo{year}{1995}).

\bibitem[{\citenamefont{Lavallard et~al.}(1996)\citenamefont{Lavallard,
  Rosenbauer, and Gacoin}}]{Lav1996}
\bibinfo{author}{\bibfnamefont{P.}~\bibnamefont{Lavallard}},
  \bibinfo{author}{\bibfnamefont{M.}~\bibnamefont{Rosenbauer}},
  \bibnamefont{and} \bibinfo{author}{\bibfnamefont{T.}~\bibnamefont{Gacoin}},
  \bibinfo{journal}{Phys. Rev A} \textbf{\bibinfo{volume}{54}},
  \bibinfo{pages}{5450} (\bibinfo{year}{1996}).

\bibitem[{\citenamefont{Lamouche et~al.}(1998)\citenamefont{Lamouche,
  Lavallard, and Gacoin}}]{Lam1998}
\bibinfo{author}{\bibfnamefont{G.}~\bibnamefont{Lamouche}},
  \bibinfo{author}{\bibfnamefont{P.}~\bibnamefont{Lavallard}},
  \bibnamefont{and} \bibinfo{author}{\bibfnamefont{T.}~\bibnamefont{Gacoin}},
  \bibinfo{journal}{J. Lumin} \textbf{\bibinfo{volume}{76\&77}},
  \bibinfo{pages}{662} (\bibinfo{year}{1998}).

\bibitem[{\citenamefont{Meltzer et~al.}(1999)\citenamefont{Meltzer, Feofilov,
  Tissue, and Yuan}}]{Mel1999}
\bibinfo{author}{\bibfnamefont{R.~S.} \bibnamefont{Meltzer}},
  \bibinfo{author}{\bibfnamefont{S.~P.} \bibnamefont{Feofilov}},
  \bibinfo{author}{\bibfnamefont{B.}~\bibnamefont{Tissue}}, \bibnamefont{and}
  \bibinfo{author}{\bibfnamefont{H.~B.} \bibnamefont{Yuan}},
  \bibinfo{journal}{Phys. Rev B} \textbf{\bibinfo{volume}{60}},
  \bibinfo{pages}{R14012} (\bibinfo{year}{1999}).

\bibitem[{\citenamefont{Duan and Reid}(2005{\natexlab{a}})}]{Dua2005d}
\bibinfo{author}{\bibfnamefont{C.~K.} \bibnamefont{Duan}} \bibnamefont{and}
  \bibinfo{author}{\bibfnamefont{M.~F.} \bibnamefont{Reid}}
  (\bibinfo{year}{2005}{\natexlab{a}}), \bibinfo{note}{to be presented on the
  24th international rare-earth research conference, Keystone, Colorado, June
  26-30}.

\bibitem[{\citenamefont{Wuister et~al.}(2004)\citenamefont{Wuister,
  de~M.~Donega, and Meijerink}}]{Wui2004}
\bibinfo{author}{\bibfnamefont{S.~F.} \bibnamefont{Wuister}},
  \bibinfo{author}{\bibfnamefont{C.}~\bibnamefont{de~M.~Donega}},
  \bibnamefont{and}
  \bibinfo{author}{\bibfnamefont{A.}~\bibnamefont{Meijerink}},
  \bibinfo{journal}{J. Chem. Phys} \textbf{\bibinfo{volume}{121}},
  \bibinfo{pages}{4310} (\bibinfo{year}{2004}).

\bibitem[{\citenamefont{Duan and Reid}(2005{\natexlab{b}})}]{Dua2005b}
\bibinfo{author}{\bibfnamefont{C.~K.} \bibnamefont{Duan}} \bibnamefont{and}
  \bibinfo{author}{\bibfnamefont{M.~F.} \bibnamefont{Reid}},
  \bibinfo{journal}{J. Chem. Phys} \textbf{\bibinfo{volume}{122}},
  \bibinfo{pages}{094714} (\bibinfo{year}{2005}{\natexlab{b}}).

\bibitem[{\citenamefont{Ushida et~al.}(2004)\citenamefont{Ushida, Iwadate,
  Hattori, Nishiyama, and et~al.}}]{Ush2004}
\bibinfo{author}{\bibfnamefont{H.}~\bibnamefont{Ushida}},
  \bibinfo{author}{\bibfnamefont{Y.}~\bibnamefont{Iwadate}},
  \bibinfo{author}{\bibfnamefont{T.}~\bibnamefont{Hattori}},
  \bibinfo{author}{\bibfnamefont{S.}~\bibnamefont{Nishiyama}},
  \bibnamefont{and} \bibinfo{author}{\bibnamefont{et~al.}},
  \bibinfo{journal}{J. Alloys Compd.} \textbf{\bibinfo{volume}{377}},
  \bibinfo{pages}{167} (\bibinfo{year}{2004}).

\end{thebibliography}

\clearpage


\end{document}